\documentclass[preprint,aps,nofootinbib]{revtex4}
\usepackage{amsfonts,amsmath,amssymb,amsthm}
\usepackage{latexsym}
\usepackage{bbm,bm}
\usepackage{graphicx}

\newcommand{\ket}[1]{\lvert #1 \rangle}
\newcommand{\bra}[1]{\langle #1 \lvert}
\newcommand{\beq}{\begin{equation}}
\newcommand{\eeq}{\end{equation}}
\newcommand{\beqs}{\begin{eqnarray}}
\newcommand{\eeqs}{\end{eqnarray}}

\begin{document}

\title{Protection of Entanglement in the presence of Markovian or Non-Markovian Environment via particle velocity : Exact Results}

\author{ DaeKil Park$^{1,2}$\footnote{dkpark@kyungnam.ac.kr}}

\affiliation{$^1$Department of Electronic Engineering, Kyungnam University, Changwon
                 631-701, Korea    \\
             $^2$Department of Physics, Kyungnam University, Changwon
                  631-701, Korea    
                      }

\begin{abstract}
On the analytic ground we examine a physical mechanism how particle velocity can protect an entanglement when quantum system is embedded in 
Markovian or non-Markovian environment. In particular the effect of particle velocity is examined in the entanglement sudden death (ESD) and revival of 
entanglement (ROE) phenomena. Even though particles move fast, the ESD phenomenon does not disappear if it occurs at zero velocity. 
However the time domain $0 \leq t \leq t_*$ for nonvanishing entanglement becomes larger and larger with increasing velocity.
When ROE phenomenon occurs at zero velocity, even small velocity can make this phenomenon not to occur although the oscillatory behavior of
entanglement in time is maintained. For comparatively large velocity the amplitude of the oscillatory behavior becomes extremely small. 
In this way the entanglement can
be protected by particle velocity. The protection of entanglement via velocity is compared with that via the detuning parameter.
\end{abstract}

\maketitle
Quantum entanglement\cite{horodecki09} is the most important physical resource for development of quantum technology.
As shown for last two decades it plays a crucial role in quantum teleportation\cite{teleportation},
superdense coding\cite{superdense}, quantum cloning\cite{clon}, and quantum cryptography\cite{cryptography}. It is also quantum entanglement, 
which makes the quantum computer outperform the classical one\cite{text,computer}. 

Since quantum entanglement is purely quantum property, it can be maintained in time only in ideally isolated system. 
However, real physical systems inevitably interact with their surroundings. Thus, physical system loses its entanglement by contacting 
the environment.  In this reason  we expect that the degradation of entanglement occurs\cite{yu02-1,simon02-1,dur04-1}.

Usually, the degradation of entanglement emerges as a form of  an exponential decay in time by successive halves. 
For particular initial states, however, the entanglement sudden death (ESD) occurs  when the entangled multipartite quantum system is embedded in
Markovian environments\cite{markovian,yu05-1,yu06-1,yu09-1}. This means that the entanglement is completely disentangled at finite times. 
This ESD phenomenon has been revealed experimentally\cite{almeida07,laurat07}. 

The dynamics of entanglement was also examined when the physical system is embedded in non-Markovian environment\cite{breuer02,bellomo07}.
It has been shown that there is a revival of entanglement (ROE) after a finite period of time of its complete disappearance. This is mainly due to 
the memory effect of the non-Markovian environment. This ROE phenomenon was shown in Ref.\cite{bellomo07} by making use 
of the two qubit system and concurrence\cite{concurrence1} as a bipartite entanglement measure. Subsequently, many works have been done 
to quantify the non-Markovianity\cite{breuer09,vacchini11,chruscinski11,rivas14,hall14,kwang15-1,park16}.

The degradation of entanglement is a crucial obstacle in real quantum information processing. In order to overcome this problem we should  
reduce the effect of decoherence as much as possible. For this purpose various techniques were developed for Markovian\cite{protection1} and 
non-Markovian\cite{protection2} environments. Recently, it was shown that the protection of 
entanglement is possible by increasing the particle velocity when the quantum system is embedded in the non-Markovian environment\cite{velocity}.
The authors in Ref.\cite{velocity} examined the effect of velocity by applying the fourth-order Runge-Kutta numerical method. However, it is in general
difficult to understand the physical mechanism exactly from a numerical technique. In order to understand how the particle velocity reduces 
the effect of decoherence we need to reconsider this issue on the analytic ground, which is main motivation of present paper.
There is another minor motivation. The authors of Ref.\cite{velocity} argued that the dynamics of entanglement is not dependent on the particle 
velocity and the transition frequency individually, but depends on their multiplication. However, we cannot find any physical reason for this dependence. 
Our analytic
approach shows that this argument is not true, but is approximately true for some cases. Thus, 
we examine again the effect of particle velocity in the presence of Markovian or non-Markovian environment analytically. In particular, we examine 
in detail how the ESD and ROE phenomena are affected by particle velocity. 

We choose exactly the same physical setup with that of Ref.\cite{velocity}, that is, the whole system is composed of two non-interacting identical 
systems. Each subsystem consists of an atom qubit and a structured environment made of two perfect reflecting mirrors at the positions 
$z = -L$ and $z = \ell$ with a partially reflecting mirror at $z =0$. The electromagnetic fields inside the cavities plays a role of environment. 

We will briefly describe how the entanglement dynamics can be derived schematically. The detailed derivation is in Ref.\cite{velocity}. 
The dynamics of one atom and its environment is governed by the  Schr\"{o}dinger equation 
$\frac{d}{dt} \psi (t) = -i H_I (t) \psi (t)$, where $H_I$ is a Hamiltonian for single atom and its interaction with an environment. 
Solving this Schr\"{o}dinger equation with appropriate boundary conditions arising in the cavities, one can derive the state $\rho (t)$ of single atom by taking a partial trace with respect to its environment, that is, $\rho (t) = \mbox{tr}_{env} \ket{\psi(t)} \bra{\psi(t)}$.
Since two atoms interacts only and independently with its own environment, the quantum state of two atoms can be derived by the Kraus 
operators\cite{kraus83}. For example, if the initial state of two atoms is 
\begin{equation}
\label{initial1}
\ket{\Psi} = a \ket{00} + b \ket{11}
\end{equation}
where $a$ is real and positive, and 
$b = |b| e^{i \delta}$ with $a^2 + |b|^2 = 1$, the concurrence at time $t$ is given by 
\begin{equation}
\label{dynamics1}
{\cal C} (t) = \max \bigg[ 0, 2 |b| |P(t)|^2 \big( a - |b| [1 - |P(t)|^2 ] \big) \bigg],
\end{equation}
where $P(t)$ satisfies the integral equation 
\begin{equation}
 \label{effect1}
 \frac{d}{d t} P(t) = - \int_0^t d s f(t, s) P(s).
 \end{equation}
 If  velocity of atom is $v$ and its transition frequency is $\omega_0$,  the correlation function $f(t, s)$ is given by 
\begin{equation}
 \label{correlation}
 f(t, s) = \int_{\infty}^{\infty} d \omega J (\omega) \sin \left[\omega (\beta t - \tau) \right] \sin \left[\omega (\beta s - \tau) \right] 
 e^{-i (\omega - \omega_0) (t - s)}
 \end{equation}
 where $\beta = v / c$, $\tau = \ell / c$, and the spectral density\cite{breuer02} is
\begin{equation}
 \label{spectral}
 J (\omega) = \frac{1}{2 \pi} \frac{\gamma_0 \lambda^2}{(\omega_0 - \omega - \Delta)^2 + \lambda^2}.
 \end{equation}
 In order to neglect the relativistic effect we should require $\beta << 1$.
 In equation (\ref{correlation}) the sine terms in integral comes from boundary conditions at the mirrors. In Eq. (\ref{spectral}) 
 the parameter $\lambda$ defines the spectral width of the coupling, which is connected to the reservoir 
correlation time $\tau_B$ by the relation $\tau_B \approx 1 / \lambda$ and the relaxation time scale $\tau_R$ on which the state of the system
changes is related to $\gamma_0$ by $\tau_R \approx 1 / \gamma_0$. The parameter $\Delta$ is a detuning parameter.
Thus,  the center frequency of the cavity is detuned by an amount $\Delta$ against the atomic transition frequency $\omega_0$. 
 
Now, we want to compute $P(t)$ analytically as much as possible.
 Inserting Eq. (\ref{spectral}) into Eq. (\ref{correlation}) $f(t, s)$ can be written as 
 \begin{equation}
 \label{correlation1}
 f(t, s) = - \frac{\gamma_0 \lambda^2}{8 \pi} \sum_{j=1}^4 f_j (t, s)
 \end{equation}
 where
 \begin{eqnarray}
 \label{f1to4}
&&  f_1 (t, s) = \exp \bigg[ i (\omega_0 - \Delta) \left\{\beta (t+s) - 2 \tau \right\} + i \Delta (t - s) \bigg] 
                \int_{-\infty}^{\infty} d u \frac{e^{-iu [(1 - \beta) t - (1 + \beta) s + 2 \tau]}}{u^2 + \lambda^2}         \nonumber   \\
&&  f_2 (t, s) = \exp \bigg[- i (\omega_0 - \Delta) \left\{\beta (t+s) - 2 \tau \right\} + i \Delta (t - s) \bigg] 
                \int_{-\infty}^{\infty} d u \frac{e^{-iu [(1 + \beta) t - (1 - \beta) s - 2 \tau]}}{u^2 + \lambda^2}         \nonumber    \\
&& f_3 (t, s) = - e^{i (t - s) [\beta \omega_0 + (1 - \beta) \Delta]} \int_{-\infty}^{\infty} d u \frac{e^{-i u (1 - \beta) (t - s)}}{u^2 + \lambda^2}
                                                                                                                                                                            \\     \nonumber
&& f_4 (t, s) = - e^{i (t - s) [-\beta \omega_0 + (1 + \beta) \Delta]} \int_{-\infty}^{\infty} d u \frac{e^{-i u (1 + \beta) (t - s)}}{u^2 + \lambda^2}.
 \end{eqnarray}
 Making use of 
 \begin{equation}
 \label{using-1}
 \int_{-\infty}^{\infty} \frac{e^{-i a u}}{u^2 + \lambda^2} d u = \frac{\pi}{\lambda} e^{-\lambda |a|},
 \end{equation}
 the correlation function $f(t,s)$ in Eq. (\ref{correlation1}) reduces to 
 \begin{eqnarray}
 \label{correlation2}
&& f(t, s) = \frac{\gamma_0 \lambda}{8} \Bigg[ e^{i [\beta \omega_0 + (1 - \beta) \Delta ] (t - s)} e^{-\lambda (1 - \beta) |t-s|} 
                                                                   +  e^{i [-\beta \omega_0 + (1 + \beta) \Delta ] (t - s)} e^{-\lambda (1 + \beta) |t-s|}  \\  \nonumber
&&  \hspace{1.8cm}
-\exp\bigg\{i (\omega_0 - \Delta) [\beta (t+s) - 2 \tau] + i \Delta (t-s) \bigg\} e^{-\lambda |(1 - \beta) t - (1 + \beta) s + 2 \tau |}  
                                                                                                                                                                                                       \\   \nonumber
&&  \hspace{1.8cm}
-\exp\bigg\{-i (\omega_0 - \Delta) [\beta (t+s) - 2 \tau] + i \Delta (t-s) \bigg\} e^{-\lambda |(1 + \beta) t - (1 - \beta) s - 2 \tau |} 
                                                            \Bigg].
\end{eqnarray}
In the continuum limit ($\tau \rightarrow \infty$) Eq. (\ref{correlation2}) is simplified as 
\begin{equation}
\label{correlation3}
f(t, s) = \frac{\gamma_0 \lambda}{8} \Bigg[ e^{i [\beta \omega_0 + (1 - \beta) \Delta ] (t - s) -\lambda (1 - \beta) |t-s|} 
                                                                   +  e^{i [-\beta \omega_0 + (1 + \beta) \Delta ] (t - s)-\lambda (1 + \beta) |t-s|} \Bigg].
\end{equation}
When $t > s$, it is more simplifies in a form
\begin{equation}
\label{correlation4}
f(t, s) = g(t - s)
\end{equation}
where
\begin{equation}
\label{defg1}
g(t) = \frac{\gamma_0 \lambda}{4} \cosh (\alpha t) e^{-\bar{\lambda} t}
\end{equation}
with $\bar{\lambda} = \lambda - i \Delta$ and $\alpha = \beta (\bar{\lambda} + i \omega_0)$.
Thus Eq. (\ref{effect1}) simply reduces to 
\begin{equation}
\label{effect2}
\frac{d}{d t} P(t) = -(g * P) (t)
\end{equation}
where $*$ means a convolution.

In order to derive $P(t)$ explicitly  we take a Laplace transform $\hat{f} (p) \equiv \int_0^{\infty} f(t) e^{-p t} d t$ to both sides of Eq. (\ref{effect2}).
Using $P(t = 0) = 1$ one can show easily 
\begin{equation}
\label{effect3}
\hat{P} (p) = \frac{1}{p + \hat{g} (p)}.
\end{equation}
From Eq. (\ref{defg1}) it is also easy to show
\begin{equation}
\label{defg2}
\hat{g} (p) = \frac{\gamma_0 \lambda}{8} \left( \frac{1} {p + v_+} + \frac{1}{p + v_-} \right)
\end{equation}
where $v_{\pm} \equiv \bar{\lambda} \pm \alpha$. Inserting Eq. (\ref{defg2}) into Eq. (\ref{effect3}) one can show directly
\begin{equation}
\label{effect4}
\hat{P} (p) = \frac{(p + v_+) (p + v_-)}{(p - p_1) (p - p_2) (p - p_3)}
\end{equation}
where $p_j \hspace{.2cm} (j=1,2,3)$ are roots of 
\begin{equation}
\label{cubic1}
p^3 + 2 \bar{\lambda} p^2 +  \left(v_+ v_- + \frac{\gamma_0 \lambda}{4} \right) p + \frac{\gamma_0 \lambda \bar{\lambda}}{4} = 0.
\end{equation}
Since general cubic equation can be solved analytically, it is always possible to obtain the analytical expressions of $p_j$ even though their expressions
are too lengthy except few special cases.

The inverse Laplace transform of Eq. (\ref{effect4}) can be easily performed by applying the Bromwich integral formula in complex plane, i.e.
\begin{equation}
\label{bromwich}
P(t) = \mbox{sum of residues of } \hat{P} (z) e^{z t}.
\end{equation}
Thus $P(t)$ becomes
\begin{equation}
\label{effect5}
P(t) = \frac{(p_1 + v_+) (p_1 + v_-)}{(p_1 - p_2) (p_1 - p_3)} e^{p_1 t} - \frac{(p_2 + v_+) (p_2 + v_-)}{(p_1 - p_2) (p_2 - p_3)} e^{p_2 t}
+ \frac{(p_3 + v_+) (p_3 + v_-)}{(p_1 - p_3) (p_2 - p_3)} e^{p_3 t}.
\end{equation}

In real calculation it is convenient to introduce following dimensionless parameters
\begin{equation}
\label{dimensionless}
x_1 = \frac{\lambda}{\gamma_0}     \hspace{1.0cm}
x_2 = \frac{\omega_0}{\gamma_0}   \hspace{1.0cm}
x_3 = \frac{\Delta}{\gamma_0}         \hspace{1.0cm}
x = \gamma_0 t.
\end{equation}
Now we define $q \equiv p / \gamma_0$. Then the cubic equation Eq. (\ref{cubic1}) reduces to 
\begin{equation}
\label{cubic2}
q^3 + 2 (x_1 - i x_3) q^2 + \left(u_+ u_- + \frac{x_1}{4} \right) q + \frac{x_1 (x_1 - i x_3)}{4} = 0
\end{equation}
where 
\begin{equation}
\label{defupm}
u_{\pm} \equiv \frac{v_{\pm}}{\gamma_0} = (1 \pm \beta) x_1 \pm i \beta x_2 - i (1 \pm \beta) x_3.
\end{equation}
If $q_j \hspace{.2cm} (j=1,2,3)$ are roots of Eq. (\ref{cubic2}), $P(t)$ can be expressed as 
\begin{equation}
\label{effect6}
P(t) = \frac{(q_1 + u_+) (q_1 + u_-)}{(q_1 - q_2) (q_1 - q_3)} e^{q_1 x} - \frac{(q_2 + u_+) (q_2 + u_-)}{(q_1 - q_2) (q_2 - q_3)} e^{q_2 x}
+ \frac{(q_3 + u_+) (q_3 + u_-)}{(q_1 - q_3) (q_2 - q_3)} e^{q_3 x}.
\end{equation}

\begin{figure}[ht!]
\begin{center}
\includegraphics[height=5.2cm]{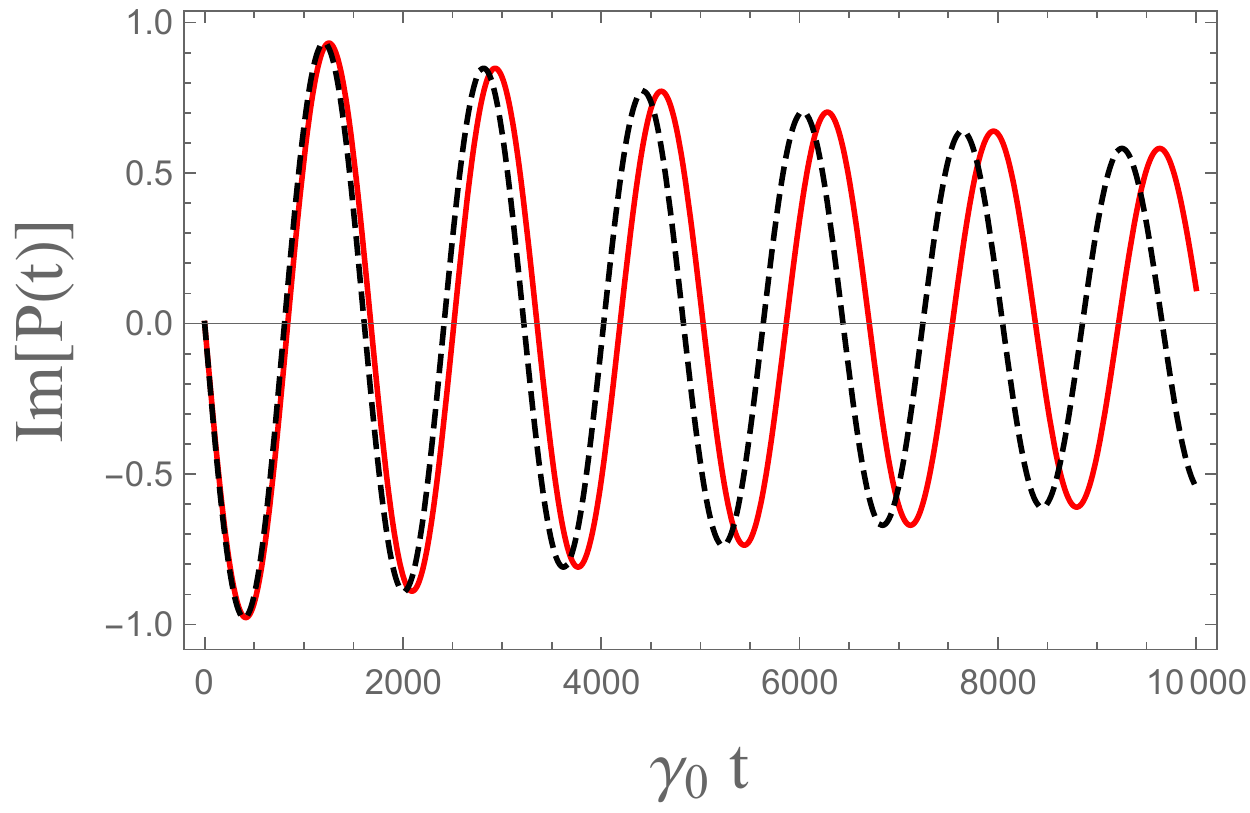}
\caption[fig1]{(Color online) The  $\gamma_0 t$-dependence of  Im$[P(t)]$ when  $\lambda = 1.5 \gamma_0$ and $\Delta = 100 \gamma_0$. The 
red solid and black dashed lines correspond to ($\beta = 0.01$, $\omega_0 = 10 \gamma_0$) and ($\beta = 0.2$, $\omega_0 = 0.5 \gamma_0$)
respectively.   The discrepancy of these two lines implies that the dynamics of entanglement depends on $\beta$ and $\omega_0$ individually.}
\end{center}
\end{figure}

As commented earlier authors in Ref.\cite{velocity} claimed that the dynamics of entanglement is not dependent on $\beta$ and $\omega_0$ individually, but  depends on $\beta \omega_0$. However, this is not correct statement. The cubic equation (\ref{cubic2}) depends on $\beta$ and $x_2$
only through $u_{\pm}$. Since $\beta << 1$, $u_{\pm}$ can be written approximately as 
$u_{\pm} \approx (x_1 - i x_3) \pm i \beta x_2$ provided that $x_2$ is comparatively larger than $x_1$ and $x_3$. In this case 
the statement of Ref.\cite{velocity} is right approximately. For other cases the dynamics of entanglement depends on $\beta$ and $\omega_0$
individually. In order to show this explicitly we plot the $\gamma_0 t$-dependence of  Im$[P(t)]$ in Fig. 1. The red solid line corresponds to 
$\beta = 0.01$ and $\omega_0 = 10 \gamma_0$, and black dashed line is for $\beta = 0.2$ and $\omega_0 = 0.5 \gamma_0$. Other parameters are 
chosen as $\lambda = 1.5 \gamma_0$ and $\Delta = 100 \gamma_0$. The discrepancy of these two lines implies that the dynamics of entanglement is 
dependent on $\beta$ and $\omega_0$ individually.

\begin{figure}[ht!]
\begin{center}
\includegraphics[height=5.2cm]{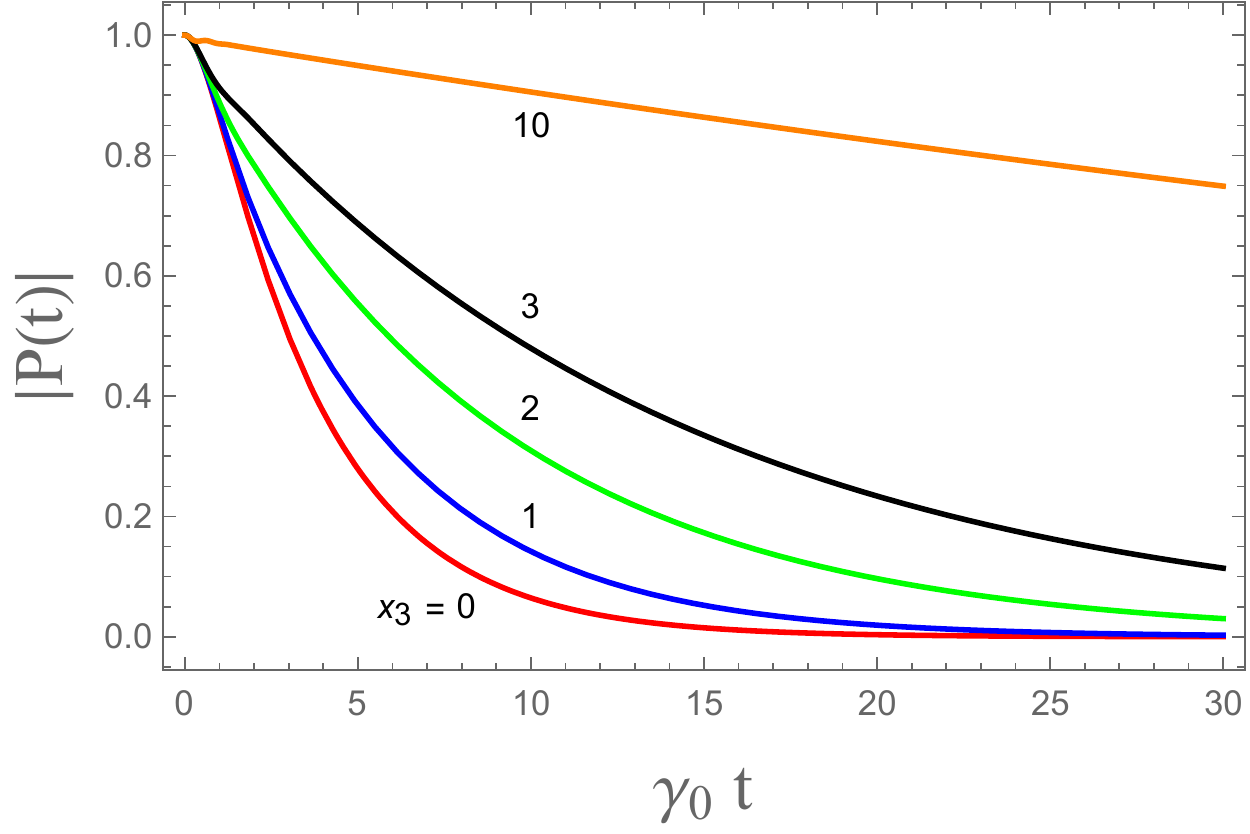}
\includegraphics[height=5.2cm]{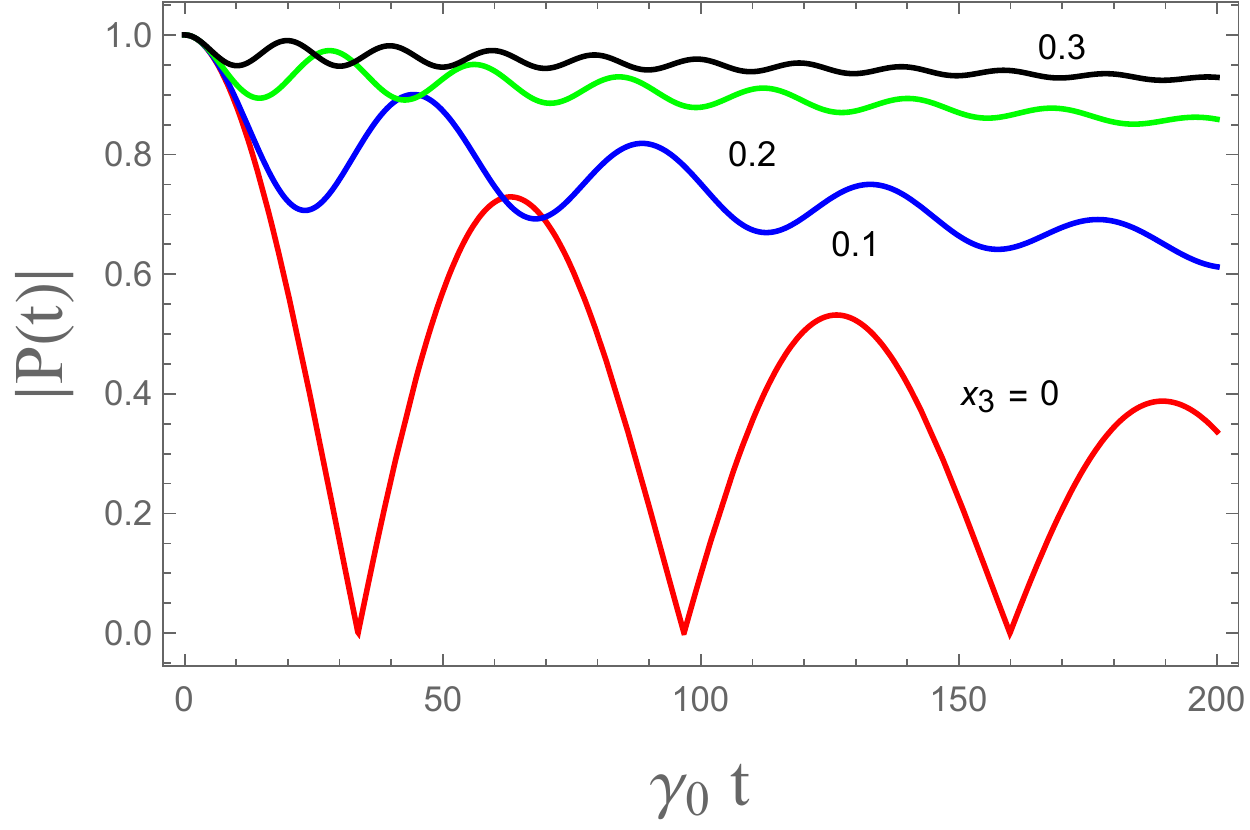}
\caption[fig2]{(Color online) Plot of $|P(t)|$ when $\beta = 0$. We choose Markovian ($x_1 = 2$) in (a) and non-Markovian ($x_1 = 0.01$)
regimes in (b) respectively with varying $\Delta$. As these figures show, the effect of decoherence is diminished with increasing $\Delta$. }
\end{center}
\end{figure}

Now, we consider few special cases. First let us consider the stationary case ($\beta = 0$). In this case $u_+$ and $u_-$ are identical as 
$u_+ = u_- \equiv u = x_1 - i x_3$. Furthermore, the roots of Eq. (\ref{cubic2}) are simply
\begin{equation}
\label{stationary1}
q_1 = -u  \hspace{.5cm}  q_2 = -\frac{1}{2} \left(u - \sqrt{u^2 - x_1} \right)   \hspace{.5cm}
 q_3 = -\frac{1}{2} \left(u + \sqrt{u^2 - x_1} \right).
 \end{equation}
Then, it is simple to derive $P(t)$ in a form
 \begin{equation}
 \label{effect7}
 P(t) = e^{-\frac{\lambda - i \Delta}{2} t} 
 \left[ \cosh \left( \frac{\sqrt{(\lambda - i \Delta)^2 - \gamma_0 \lambda}}{2} t \right) 
 + \frac{\lambda - i \Delta} {\sqrt{(\lambda - i \Delta)^2 - \gamma_0 \lambda}} \sinh  \left( \frac{\sqrt{(\lambda - i \Delta)^2 - \gamma_0 \lambda}}{2} t \right) \right].
 \end{equation}
 If $\Delta = 0$, Eq. (\ref{effect7}) reproduces the well-known expressions, that is 
 \begin{equation}
 \label{markov1}
 P(t) =  e^{-\lambda t / 2} \left[ \cosh \left(\frac{\bar{d}}{2} t \right) + \frac{\lambda}{\bar{d}} \sinh \left(\frac{\bar{d}}{2} t \right) \right]
 \end{equation}
 in the weak coupling regime $\lambda > \gamma_0$ and 
\begin{equation}
 \label{nonmarkov1}
 P(t) =  e^{-\lambda t / 2} \left[ \cos \left(\frac{d}{2} t \right) + \frac{\lambda}{d} \sin \left(\frac{d}{2} t \right) \right]
 \end{equation} 
in strong coupling regime $\lambda < \gamma_0$, where $\bar{d} = \sqrt{\lambda^2 - \gamma_0 \lambda}$ and 
$d = \sqrt{\gamma_0 \lambda - \lambda^2}$. Eq. (\ref{markov1}) and Eq. (\ref{nonmarkov1}) are responsible for the decoherence of 
Markovian and non-Markovian environments. When $\Delta \neq 0$, $P(t)$ in Eq. (\ref{effect7}) is a complex quantity. In order to explore the 
effect of $\Delta$ we plot $|P(t)|$ for $\beta = 0$ in Fig.2 with choosing $\lambda = 2 \gamma_0$ in Fig. 2(a) and $\lambda = 0.01 \gamma_0$ in Fig. 2(b). 
We also choose various $\Delta$ in each figure. As Fig. 2 exhibits, the effect of Markovian and non-Markovian environments is diminished with 
increasing $\Delta$. In this way one can protect the entanglement by making use of the detuning parameter $\Delta$\cite{protection2} too.
These figures show that non-Markovian environment is more sensitive to $\Delta$ than Markovian environment.

Another special case we consider is a slow moving case ($\beta \rightarrow 0$). In this case the roots of the cubic equation (\ref{cubic2}) can be 
obtained perturbatively as follows;
\begin{eqnarray}
\label{slow1}
&&q_1 = - u + \delta q_1 \beta^2 + {\cal O} (\beta^4)    \hspace{1.0cm}
q_2 = -\frac{1}{2} (u - \sqrt{u^2 - x_1}) + \delta q_2 \beta^2  + {\cal O} (\beta^4)    \\    \nonumber
&&     \hspace{3.5cm}
q_3 = -\frac{1}{2} (u + \sqrt{u^2 - x_1}) + \delta q_3 \beta^2  + {\cal O} (\beta^4)
\end{eqnarray}
where
\begin{eqnarray}
\label{slow2}
&&\delta q_1 = - \frac{4 u (u + i x_2)^2}{x_1}        \hspace{1.0cm}
\delta q_2 = - \frac{(u + i x_2)^2}{x_1 \sqrt{u^2 - x_1}} \left(u - \sqrt{u^2 - x_1} \right)^2   \\   \nonumber
&& \hspace{2.5cm}
\delta q_3 =  \frac{(u + i x_2)^2}{x_1 \sqrt{u^2 - x_1}} \left(u + \sqrt{u^2 - x_1} \right)^2.
\end{eqnarray}
Inserting Eq. (\ref{slow1}) into Eq. (\ref{effect6}) one can derive $P(t)$ analytically up to order of $\beta^2$.

Now, let us examine the dynamics of entanglement in the presence of Markovian or non-Markovian environment when the initial state is 
$\ket{\Psi}$ in Eq. (\ref{initial1}). Thus the bipartite entanglement at time $t$ is given by Eq. (\ref{dynamics1}). It is known that 
the entanglement is protected by not only $\Delta$ but also $\beta$ independently. We will examine how ESD and ROE phenomena are affected by
$\beta$ or $\Delta$.

\begin{figure}[ht!]
\begin{center}
\includegraphics[height=5.2cm]{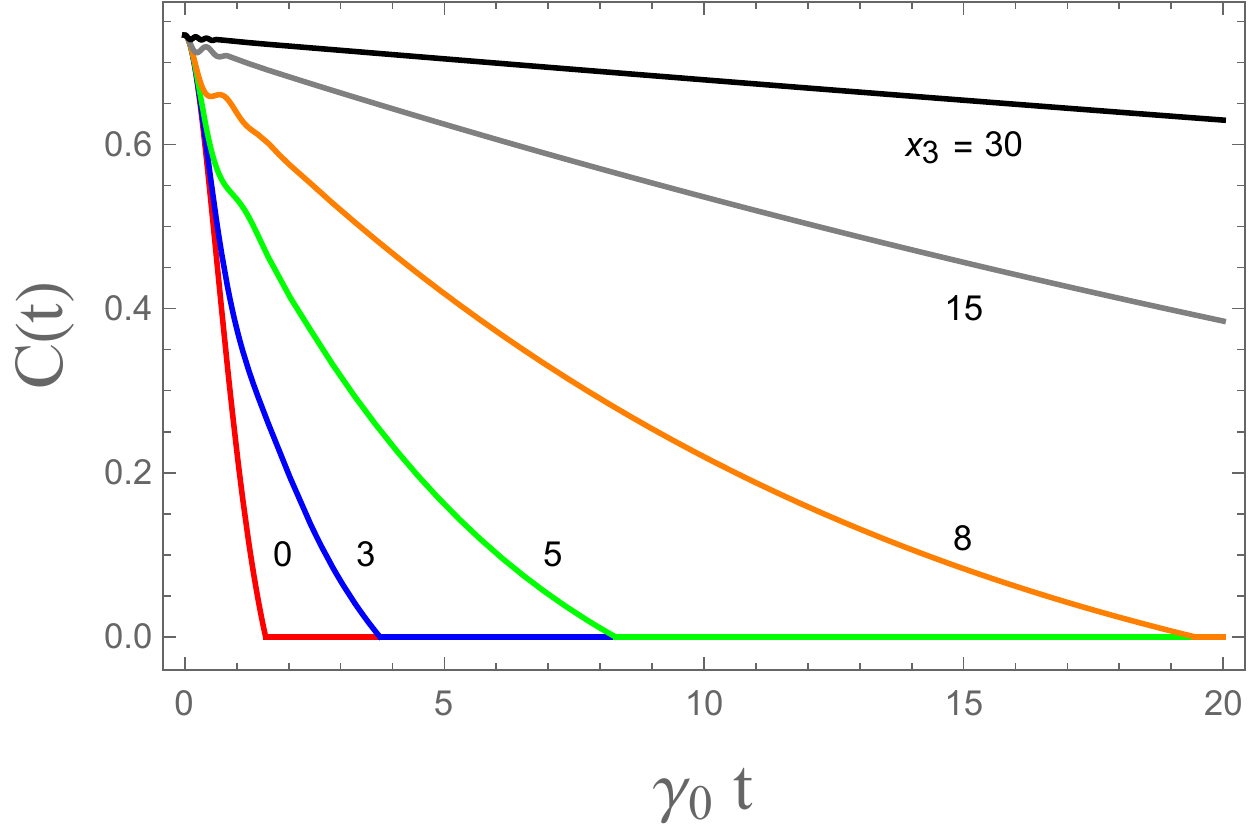}
\includegraphics[height=5.2cm]{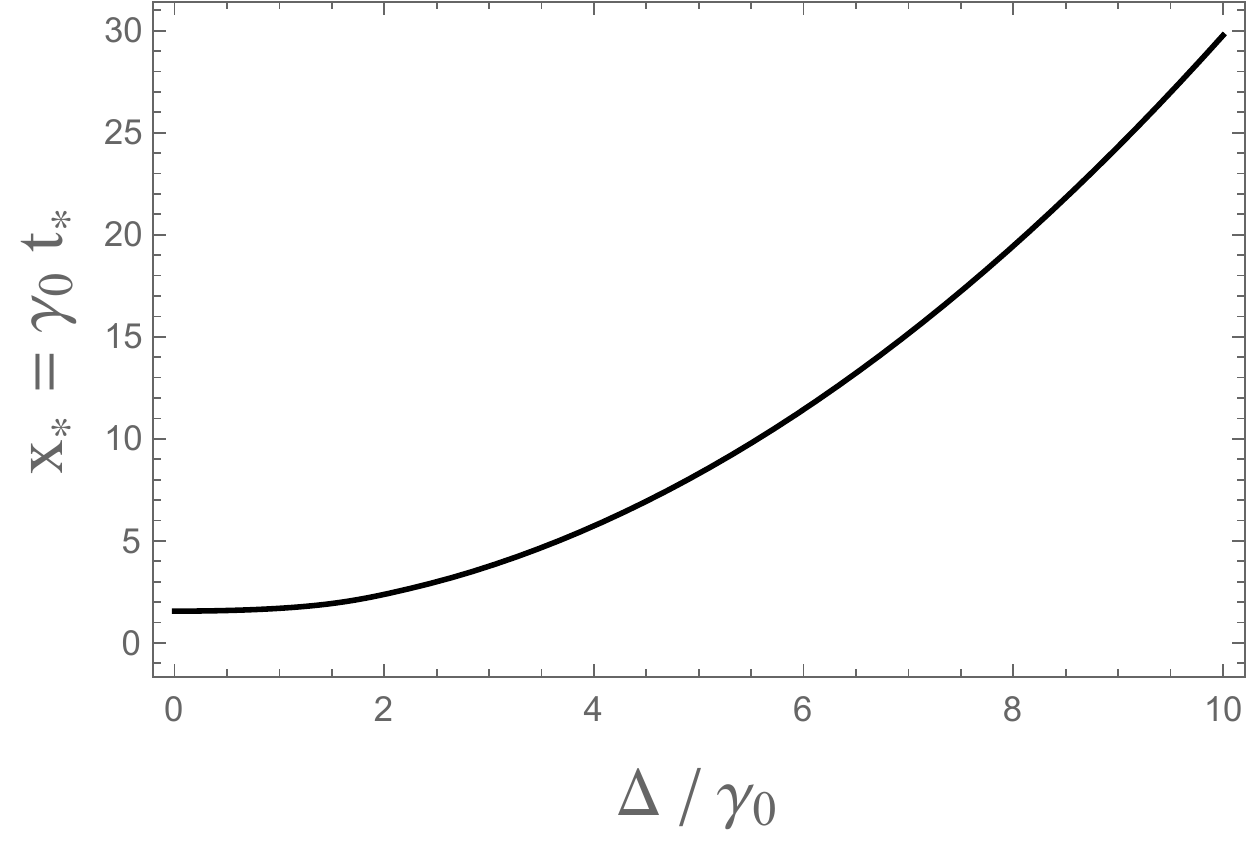}
\caption[fig3]{(Color online) (a) Plot of  ${\cal C} (t)$  with choosing  $x_3 = 0, 3, 5, 8, 15, 30$ when other parameters are fixed as 
$a = 0.4$, $\lambda = 2 \gamma_0$, and $\beta = \omega_0 = 0$. The time domain $0 \leq x \leq x_* \equiv \gamma_0 t_*$ for nonvanishing entanglement becomes larger and larger with increasing $\Delta$. (b) Plot of $x_3$-dependence of $x_*$ when same values of 
other parameters are chosen. As expected, $x_*$ increases with increasing $x_3$ quadratically. (see Eq. (\ref{fit1}))}
\end{center}
\end{figure}

In Fig. 3  we examine the effect of the detuning parameter $\Delta$ in the ESD phenomenon in the Markovian regime. 
In Fig. 3(a) we plot $\gamma_0 t$-dependence of ${\cal C} (t)$ with varying $\Delta$ when other parameters are fixed as 
$a = 0.4$, $\lambda = 2 \gamma_0$, and $\beta = \omega_0 = 0$. As this figure exhibits, the ESD phenomenon occurs regardless of $\Delta$.
However,  the time domain $0 \leq x \leq x_* \equiv \gamma_0 t_*$, 
where the entanglement is nonvanishing, becomes larger with increasing $\Delta$. In this way the entanglement is protected even in the 
Markovian environment with increasing the detuning parameter $\Delta$. In Fig. 3(b) the $x_3$-dependence of $x_*$ is plotted. As expected, $x_*$ 
increases with increasing $\Delta$. This monotonically increasing curve can be fitted as 
\begin{equation}
\label{fit1}
\gamma_0 t_* \approx 0.295318 (\Delta / \gamma_0)^2  - 0.121054 (\Delta / \gamma_0) +  1.50031.
\end{equation}

\begin{figure}[ht!]
\begin{center}
\includegraphics[height=5.2cm]{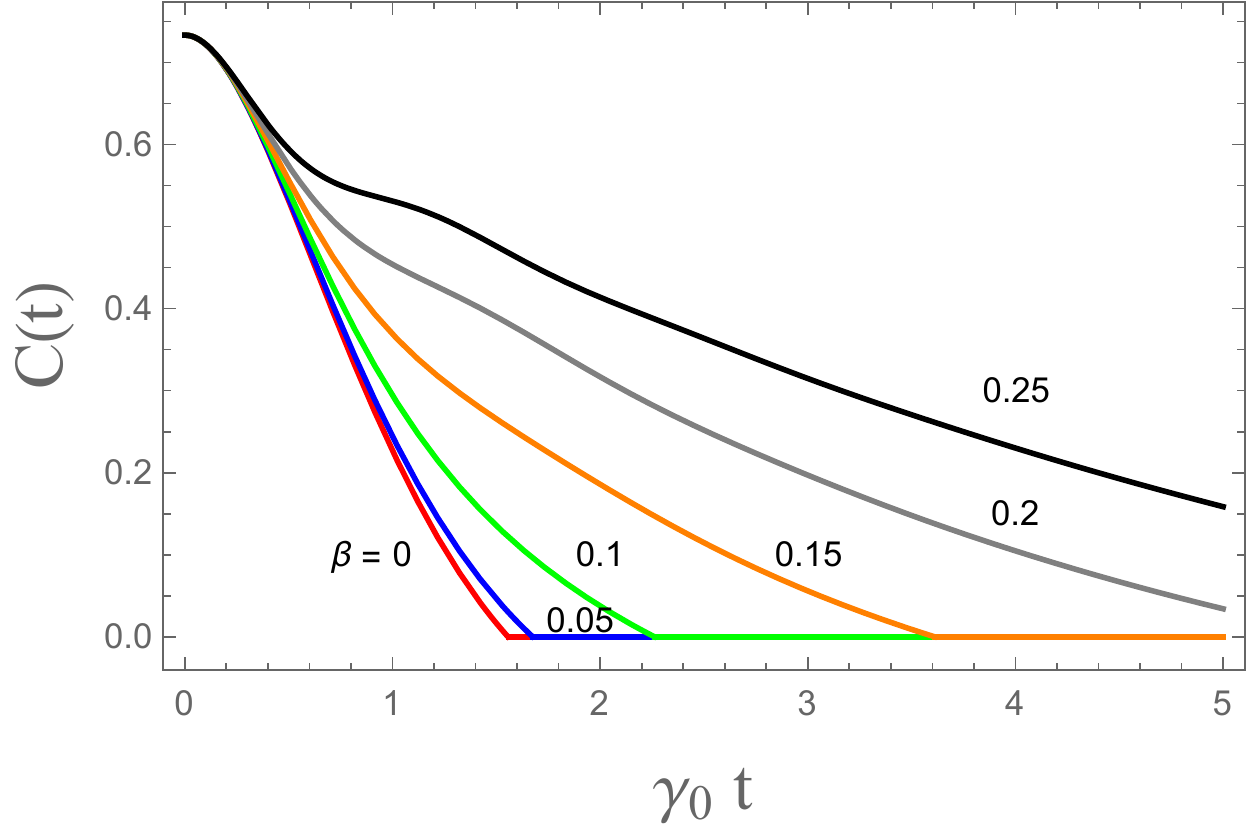}
\includegraphics[height=5.2cm]{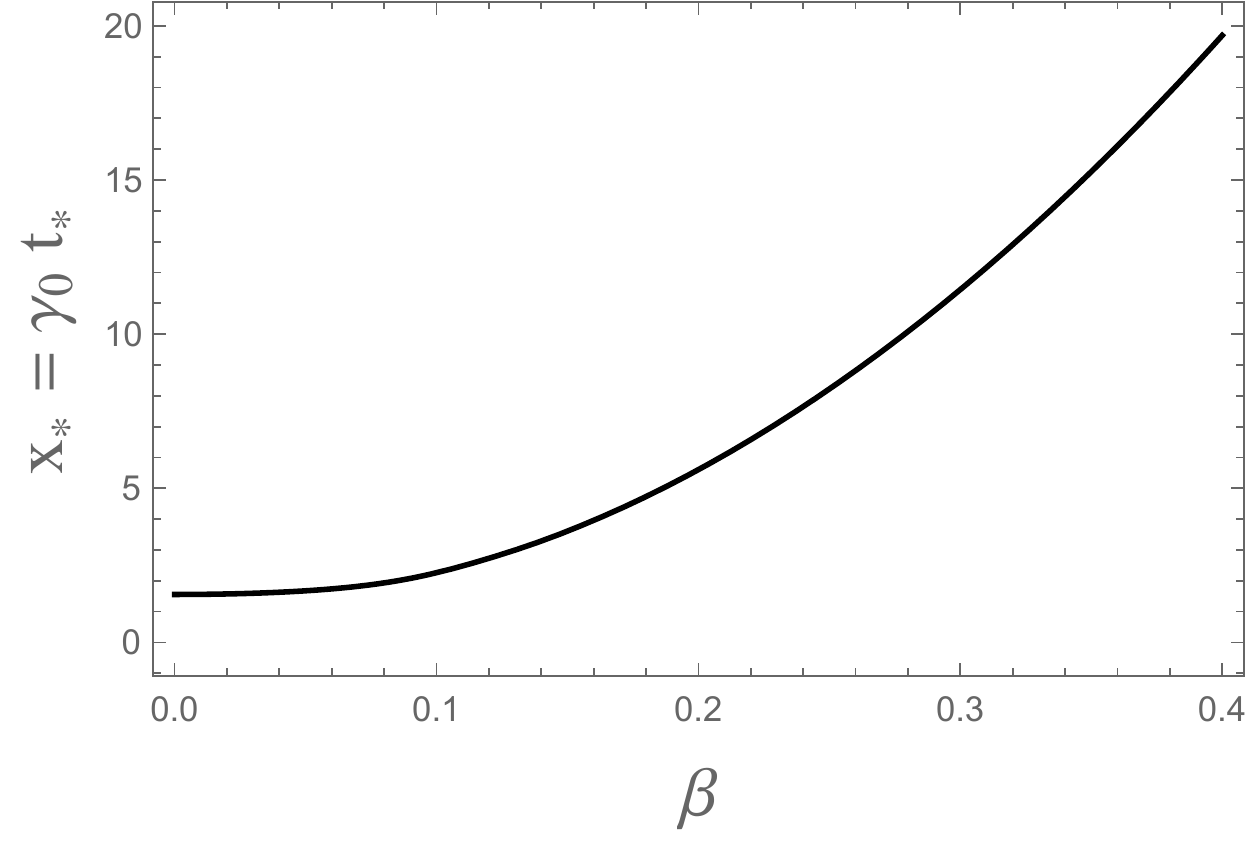}
\caption[fig4]{(Color online) (a) Plot of  ${\cal C} (t)$  with choosing  $\beta = 0, 0.05, 0.1, 0.15, 0.2, 0.25$ when other parameters are fixed as 
$a = 0.4$, $\lambda = 2 \gamma_0$, $\omega_0 = 20 \gamma_0$, and $\Delta= 0$. The time domain $0 \leq x \leq x_* \equiv \gamma_0 t_*$ for nonvanishing entanglement becomes larger and larger with increasing $\beta$. (b) Plot of $\beta$-dependence of $x_*$ when same values of 
other parameters are chosen. As expected, $x_*$ increases with increasing $\beta$ quadratically. (see Eq. (\ref{fit2})}
\end{center}
\end{figure}

The effect of the particle velocity $\beta$ on the ESD phenomenon is examined in Fig. 4. It is worthwhile noting that 
the $\beta$-dependence in cubic equation (\ref{cubic2}) is only through $u_{\pm}$ given in Eq. (\ref{defupm}). Since we will choose 
$x_1 = 2$ and $x_3 = 0$ in Fig. 3 for introducing Markovian environment and removing the effect of $\Delta$, we choose $x_2 = 20$ for 
considerable change of $u_{\pm}$. In Fig. 4(a) we plot $\gamma_0 t$-dependence of ${\cal C} (t)$ with varying $\beta$ when other 
parameters are fixed as $a = 0.4$, $\lambda = 2 \gamma_0$, $\omega_0 = 20 \gamma_0$, and $\Delta = 0$. As this figure exhibits, ESD 
phenomenon occurs regardless of $\beta$ even though the time domain $0 \leq x \leq x_* \equiv \gamma_0 t_*$, where the bipartite 
entanglement is alive, becomes larger with increasing $\beta$. This is why the entanglement can be protected in the Markovian environment by 
making use of $\beta$. In Fig. 4(b) the $\beta$-dependence of $x_*$ is plotted. This curve can be fitted as 
\begin{equation}
\label{fit2}
\gamma_0 t_* \approx 126.638 \beta^2 - 5.21879 \beta + 1.57892.
\end{equation}

\begin{figure}[ht!]
\begin{center}
\includegraphics[height=5.2cm]{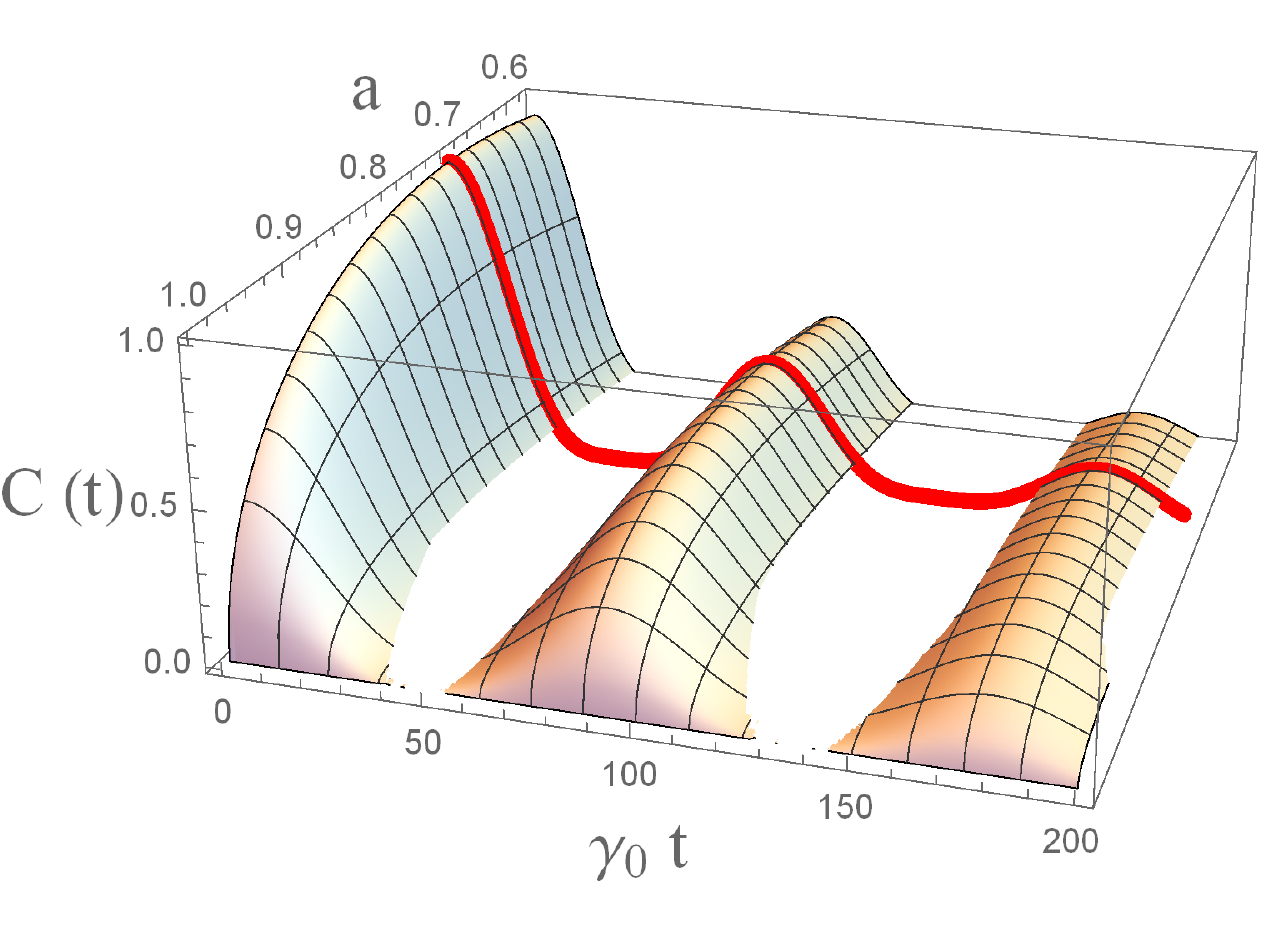}
\includegraphics[height=5.2cm]{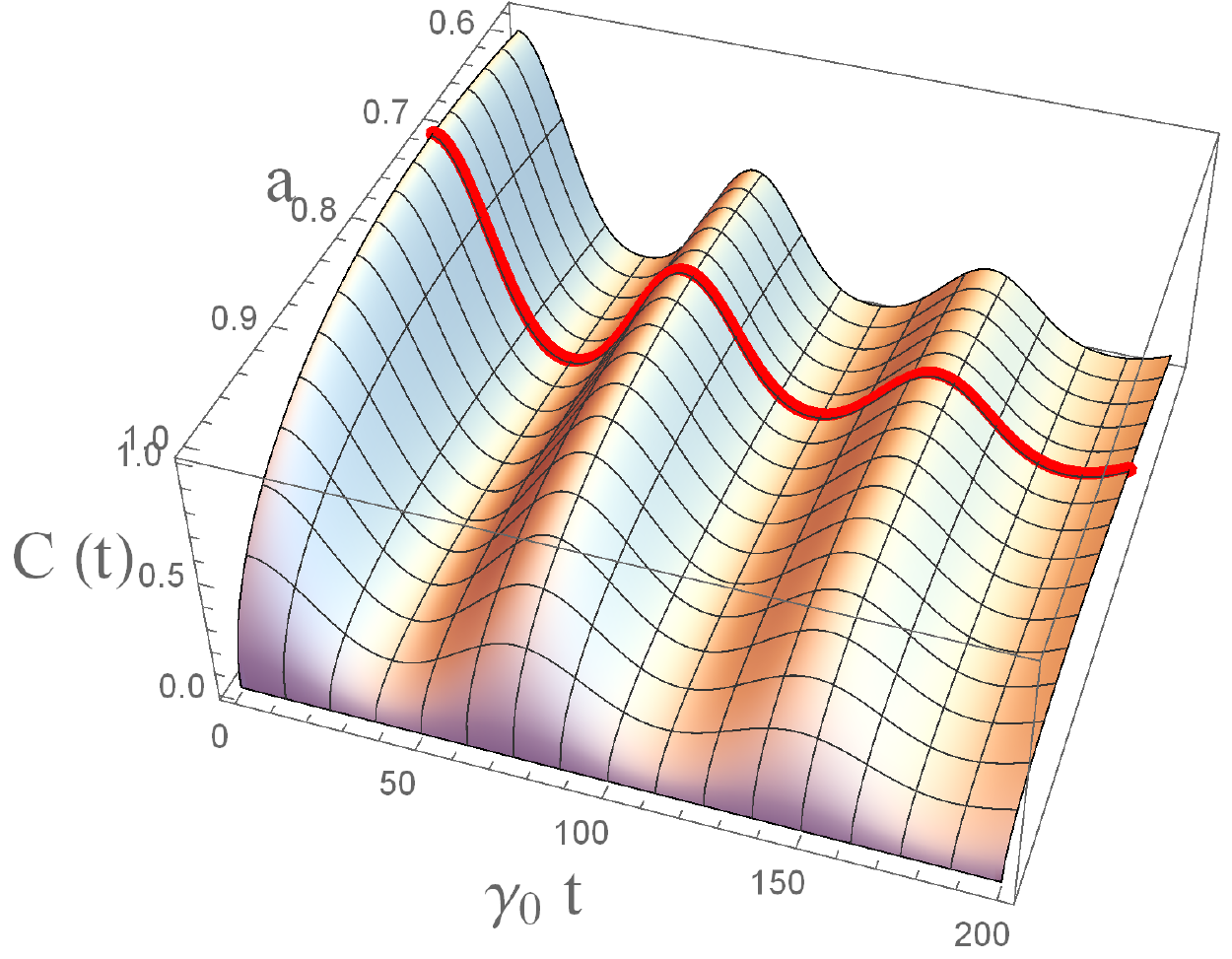}
\includegraphics[height=5.2cm]{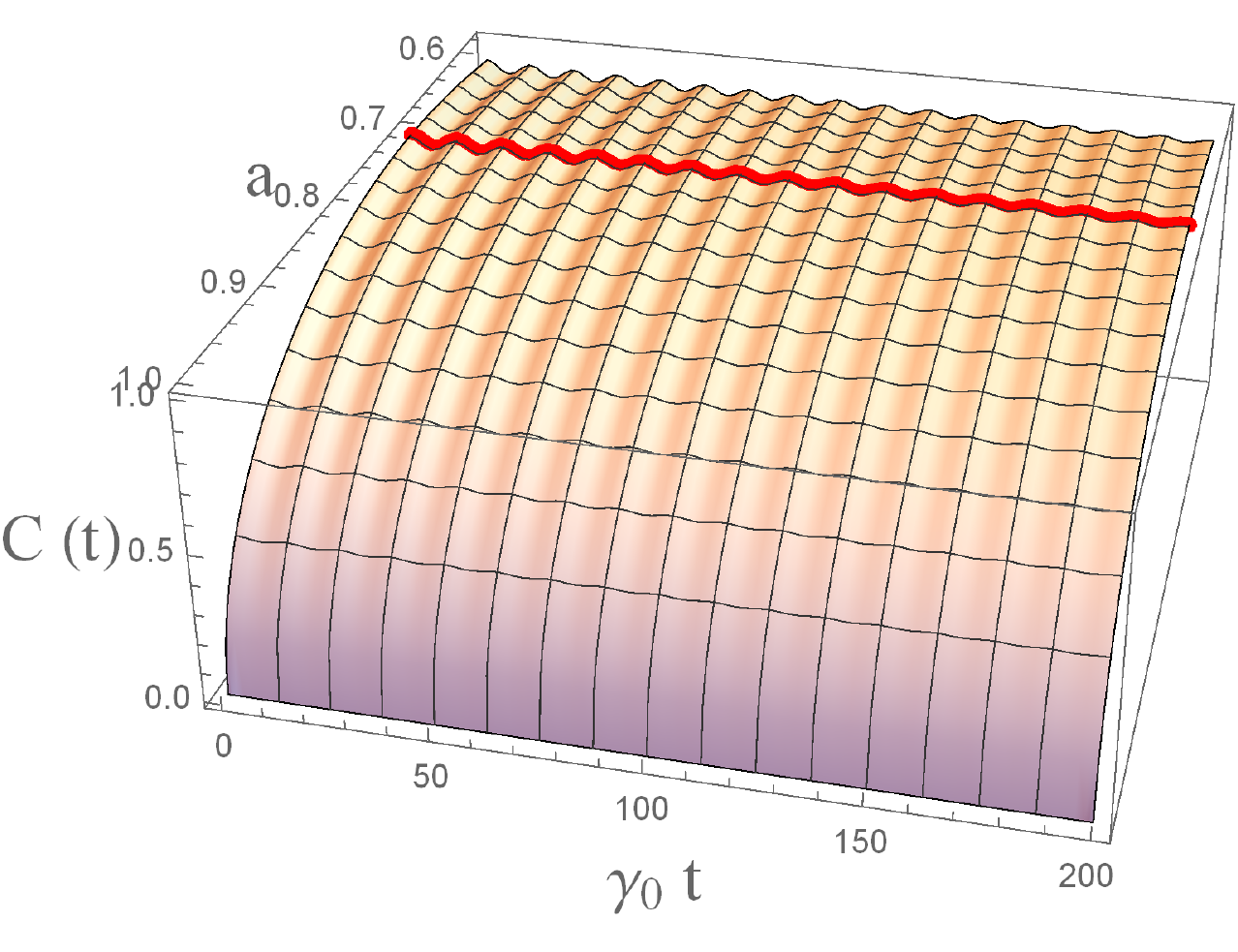}
\caption[fig5]{(Color online) We plot $(a, \gamma_0 t)$-dependence of ${\cal C}(t)$ when $\lambda = 0.005 \gamma_0$,  
$\beta = 0$, and (a) $\Delta = 0$, (b) $\Delta = 0.05 \gamma_0$, (c) $\Delta = 0.5 \gamma_0$. Fig. 5(a) exhibits a ROE phenomenon evidently.
However, this phenomenon disappears in Fig. 5(b). This means that the small increase of $\Delta$ lifts a minimum of  ${\cal C}(t)$ slightly from zero. 
Fig. 5(c) exhibits a rapid oscillatory behavior of $C (t)$ whose amplitude is very small.  
As a result, the entanglement does not decrease with the lapse of time. In this way the entanglement can be protected in the non-Markovian 
environment by making use of the detuning parameter $\Delta$.}
\end{center}
\end{figure}

Now, we examine the effect of $\beta$ and $\Delta$ on the ROE phenomenon in non-Markovian environment. First, we consider 
the effect of $\Delta$ in Fig. 5. In Fig. 5(a) we plot $(a, \gamma_0 t)$-dependence of ${\cal C}(t)$ when $\lambda = 0.005 \gamma_0$ and 
$\beta = \Delta = 0$. The line is correspondent to maximal entanglement initial state, that is, $a = 1 / \sqrt{2}$. The disconnection of wiggles in Fig. 5(a)
shows the ER phenomenon evidently. In Fig. 5(b) we increase $\Delta$ slightly as $\Delta = 0.05 \gamma_0$ without change of other parameters.
Although there are wiggles like Fig. 5(a), the wiggles in Fig. 5(b) are connected to each other. This means that the ROE phenomenon does not occur. In Fig. 5(c) we increase $\Delta$ again as $\Delta = 0.5 \gamma_0$. This figure shows
many wiggles, whose amplitude is very small. As a result, the entanglement does not decrease with the lapse of time.
In this way the entanglement can be protected in the presence of non-Markovian environment by making use of the detuning parameter $\Delta$.

\begin{figure}[ht!]
\begin{center}
\includegraphics[height=5.2cm]{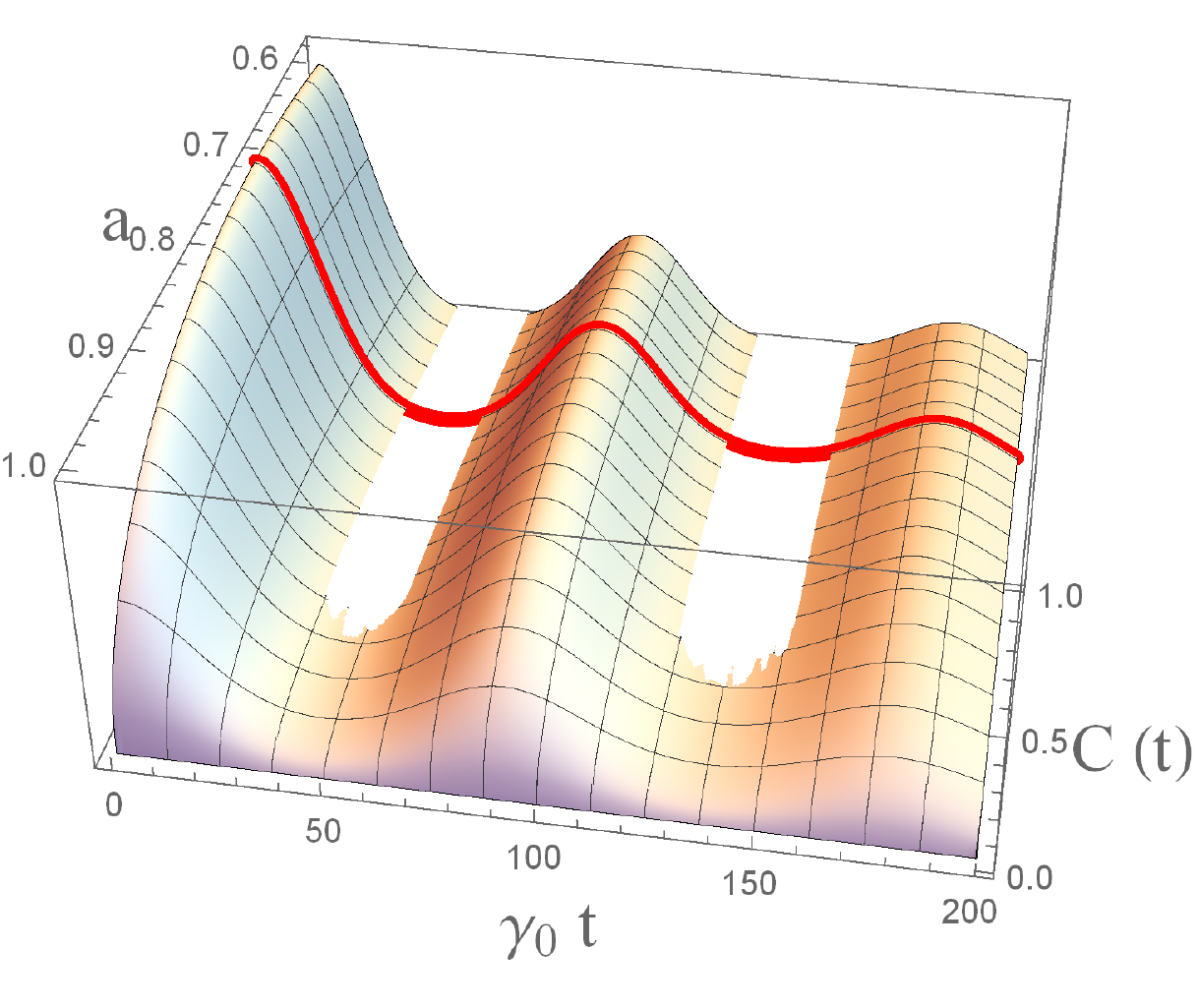}
\includegraphics[height=5.2cm]{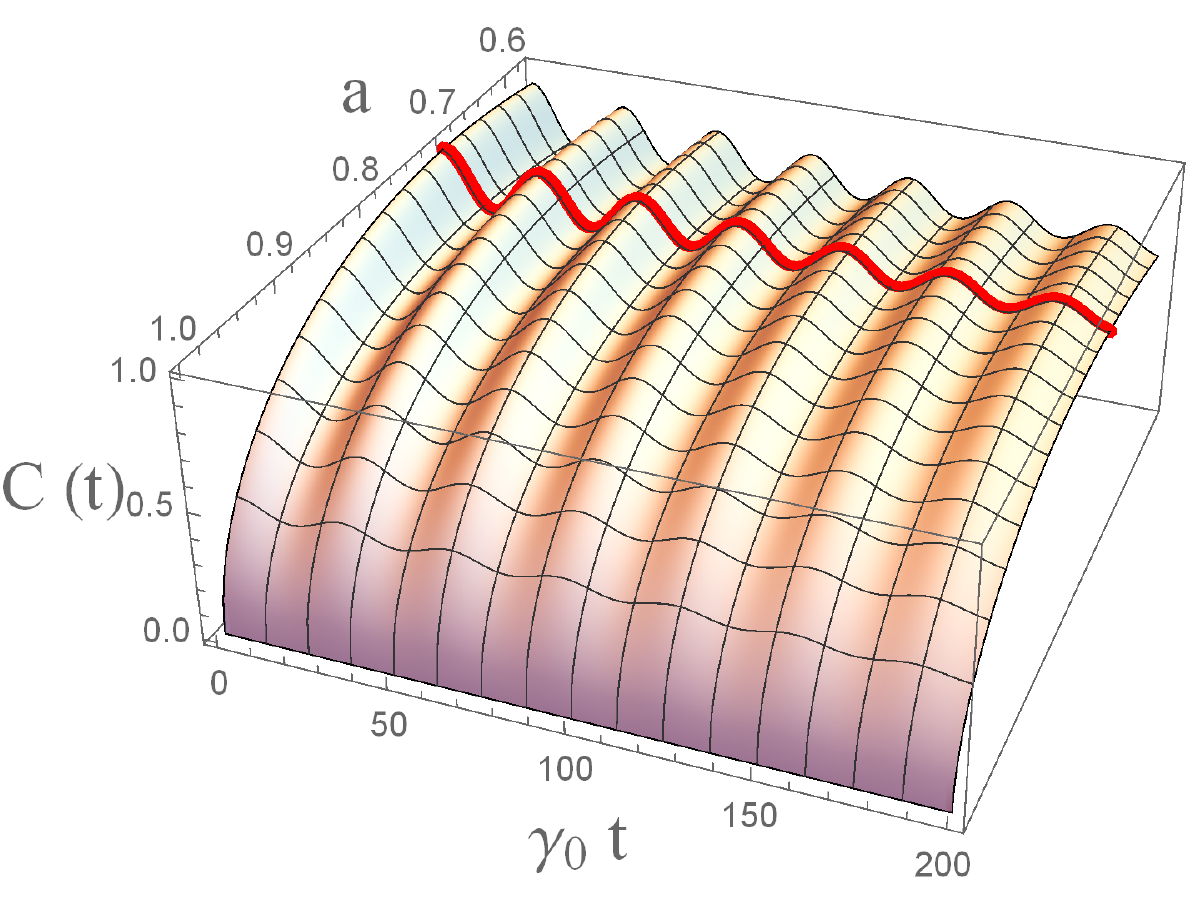}
\caption[fig6]{(Color online) We plot $(a, \gamma_0 t)$-dependence of ${\cal C}(t)$ when $\lambda = 0.005 \gamma_0$,  
$\omega_0 = 20 \gamma_0$, $\Delta = 0$, and (a) $\beta  = 0.003$, (b) $\beta = 0.01$. When $\beta = 0$, Fig. 5(a) is reproduced. 
As Fig. 6(a) exhibits, the ROE phenomenon occurs at most range of $a$ because $\beta$ is very small. Even in this case, however, the ROE
phenomenon disappears at the large $a$ region. In Fig. 6(b) the ROE phenomenon does not occur in the full range of $a$. 
Similar to Fig. 5(c) the amplitude of an oscillatory behavior of ${\cal C} (t)$ becomes small. This makes a protection of entanglement even in the 
presence of the non-Markovian environment.}
\end{center}
\end{figure}

The effect of the particle velocity in the ROE phenomenon is examined in Fig. 6. When $\lambda = 0.005 \gamma_0$,  $\beta = \Delta = 0$, and 
$\omega_0 = 20 \gamma_0$, the  $(a, \gamma_0 t)$-dependence of ${\cal C}(t)$ is exactly the same with Fig. 5(a). This is because of the fact 
that the cubic equation (\ref{cubic2}) is independent of $x_2$ when $\beta = 0$. If we change $\beta$ slightly as $\beta = 0.003$, 
 the  $(a, \gamma_0 t)$-dependence of ${\cal C}(t)$ is changed into Fig. 6(a). As this figure shows, the 
 ROE phenomenon does occur at most range of $a$. Even in this case, however, the ROE phenomenon disappears at the large $a$ region 
 due to the small increment of $\beta$. If we increase $\beta$ to 
 $0.01$,  the  $(a, \gamma_0 t)$-dependence of ${\cal C}(t)$ becomes Fig. 6(b). The ROE phenomenon does not occur in the full range of $a$. 
 Similar to Fig. 5(c) the amplitude of an oscillatory behavior of ${\cal C} (t)$ becomes small. This makes the reduction of decoherence effect in the 
 non-Markovian environment.
 
 In this paper we explore analytically the effect of particle velocity $\beta = v / c$ and detuning parameter $\Delta$ in the entanglement 
 dynamics when Markovian or non-Markovian environment is present. In particular, we examine the ESD  and ROE phenomena in the 
 Markovian and non-Markovian regimes, respectively. As Fig. 3 and Fig. 4 show, the ESD phenomenon always occurs even when $\beta$ or 
 $\Delta$ is nonzero. The difference from a case of $\beta = 0$ or $\Delta = 0$ is that the time region $0 \leq \gamma_0 t \leq \gamma_0 t_*$
 for nonvanishing entanglement becomes wider when $\beta$ or $\Delta$ becomes larger. The $\Delta$- and $\beta$-dependence of 
 $\gamma_0 t_*$ are plotted in Fig. 3(b) and Fig. 4(b). Roughly speaking, this region increases quadratically as a function of 
 $\Delta / \gamma_0$ or $\beta$.
 
 The ROE phenomenon in the non-Markovian environment is examined in Fig. 5 and Fig. 6. As these figures show, the ROE phenomenon appearing 
 in $\beta = 0$ or $\Delta = 0$ (see Fig. 5(a)) disappears for nonzero $\beta$ or nonzero $\Delta$. If $\beta$ or $\Delta$ increases more and more, 
 the amplitude of oscillatory behavior of ${\cal C} (t)$ becomes smaller and smaller in the time domain. In this way, the initial entanglement is not 
 reduced rapidly even in the presence of the non-Markovian environment.
 
 In this paper we consider only the continuum limit ($\tau = \ell / c \rightarrow \infty$). In the real physical setting, however, this limit is only
 approximation. Presumably, the effect of $\beta$ or $\Delta$ is more drastic for finite $\tau$. In this case, however, the analytic calculation 
 seems to be impossible because the convolution theorem used in this paper cannot be applied.





\end{document}